\documentclass[aps,prl,showpacs,amsmath,amssymb, 10pt]{revtex4}
\usepackage{graphicx,color}
\usepackage{amsmath}
\usepackage[font=scriptsize]{caption}
\usepackage{subcaption}
\usepackage{float}
\usepackage{color}
\usepackage{amssymb}
\usepackage{bbm}

\usepackage{amsfonts}
\usepackage{bm}

\newcommand{\eps}{\varepsilon}

\newcommand{\tr}{\mathrm{Tr}}
\newcommand{\bra}[1]{\mbox{$\langle #1 |$}}

\newcommand{\ket}[1]{\mbox{$| #1 \rangle$}}

\def\<{\langle}
\def\>{\rangle}

\newcommand{\beq}{\begin{equation}}
\newcommand{\eeq}{\end{equation}}
\newcommand{\bear}{\begin{eqnarray}}
\newcommand{\ear}{\end{eqnarray}}
\newcommand{\bdm}{\begin{displaymath}}
\newcommand{\edm}{\end{displaymath}}

\newcommand{\sig}{\sigma}
\newcommand{\ssig}{\tilde{\sigma}}

\newcommand{\al}{\alpha}
\newcommand{\be}{\beta}
\newcommand{\La}{\Lambda}
\newcommand{\g}{\gamma}

\begin{document}
\title{\textbf{Robustness and fragility of Markovian dynamics in a qubit dephasing channel}
}
\author{Filip A. Wudarski} 
\email{wudarskif@ukzn.ac.za}
\author{Francesco Petruccione}
\email{petruccione@ukzn.ac.za}
\affiliation{Quantum Research Group, School of Chemistry and Physics,
University of KwaZulu-Natal, Durban 4001, South Africa,
and National Institute for Theoretical Physics (NITheP), KwaZulu-Natal, South Africa}

\begin{abstract}
The Markovian dynamics of a qubit is investigated in the scheme of random unitary dynamics, where Kraus operators are changed by an extra noise that models imperfect experimental equipment. The behavior of Markovianity is explored in the perturbed scenario. We provide an algorithm for checking CP-divisibility (Markovianity) of a dynamical map. 
\end{abstract}
\maketitle


\section{Introduction}All quantum systems inevitably interact with the surrounding environment. In general, the access to that interaction is highly limited, because of a plethora of uncontrolled degrees of freedom. Therefore,  the theory of open quantum systems (OQS) \cite{Breuer,Weiss,Alicki} is exploited to look at the reduced dynamics of the total system and investigate the subsystem of interest.

The dynamics of OQS takes into account the influence of the environment and encodes it in completely positive and trace-preserving (CPTP) maps that are usually expressed by the Kraus representation
\begin{equation}
\La_t(\rho)=\sum_k A_k(t)\rho A_k^\dag(t),\quad \sum_k A_k^\dag(t)A_k=\mathbbm{1},
\end{equation}
where $A_k(t)$ are Kraus operators \cite{Kraus}. An appropriate knowledge of the dynamical maps gives insights into the methods of controlling and preserving fragile quantum features such as coherences and entanglement, which are a major ingredient of modern quantum technologies. 

The standard approach to OQS dynamics is based on local-in-time master equation
\begin{equation}
\dot{\La}_t=L_t\La_t,
\end{equation}
with $L_t$ being a time-dependent generator of a CPTP map $\La_t$. If the generator is time-independent, it yields a Markovian semigroup dynamics
\begin{equation}
\La_t=e^{tL},\quad \La_{t+s}=\La_t\La_s.
\end{equation}
Moreover, to have a CPTP map, the generator has to be of the celebrated Gorini-Kossakowski-Sudarshan-Lindblad (GKSL) form \cite{GKS, Lindblad}
\begin{equation}
L(\rho)=-i[H,\rho]+\sum_k \g_k\Big(V_k\rho V_k^\dag-\frac{1}{2}\{V_k^\dag V_k,\rho\}\Big),\quad H=H^\dag,\quad \g_k\ge0.
\end{equation}
The semigroup approach to OQS was the first step for investigation of Markovian and non-Markovian dynamics in the quantum regime. Nowadays, these fundamental concepts are at the core of interest, because of its vast applications, especially in quantum information theory \cite{Nilsen}. In contrast to the classical theory of stochastic processes, in the quantum case, the definitions of (non-)Markovianity are not unique and several not necessary equivalent approaches exist in the literature (see the reviews \cite{rev1,rev2}). The two most popular are the one based on the distinguishability of the quantum states \cite{BLP} and on the CP-divisibility of the dynamical map \cite{wolf1, RHP}, recently extended to the whole hierarchy of $k$-divisible maps in \cite{k-mark} (for other definitions of Markovianity see also \cite{Lu, Rajagopal, Luo, Jiang, Bogna, lorenzo, Dhar}). In this manuscript, we use the CP-divisiblity  as a definition of (non-)Markovianity, which is more restrictive than the one based on the state distinguishability. 

The evolution given by a dynamical map $\La_t$ is Markovian if and only if it can be represented as $\La_t=V_{t,s}\La_s$ $(t\ge s)$ with $V_{t,s}$ being completely positive (CP). This type of dynamics is called CP-divisible and we can always express $V_{t,s}=\La_t^{-1}\La_s$ provided  $\La_t^{-1}$  exists. In principle, finding the inverse of a dynamical map is challenging. Therefore we propose a method based on the transfer matrix approach, that helps us to decide whether $V_{t,s}$ is CP or not.  

In this research we are interested in the model of an imperfect OQS dynamics that stems from perturbation of experimental equipment. We address the following question. How robust (fragile) is Markovian dynamics while changing (time-independent) Kraus operators and keeping time factors fixed? Here we assume, that each Kraus operator can be constructed in the laboratory and be a building block for experimental simulation of OQS dynamics. 

We investigate the model based on a qubit random unitary dynamics \cite{fi1}
\begin{equation}\label{map}
\La_t(\rho)=\sum_{k=0}^3p_k(t)\sig_k\rho\sig_k^\dag,
\end{equation} 
where $\sig_k$ are Pauli matrices ($\sig_0=\mathbbm{1}$) and $\{p_k(t)\}$ is a probability distribution ($\sum p_k(t)=1$ and $p_k(t)\ge0$ for $t\ge0$). Equation (\ref{map}) is generated by a time-local generator
\begin{equation}
L_t(\rho)=\sum_{k=1}^3\g_k(t)\Big(\sig_k\rho\sig_k-\rho\Big).
\end{equation}
We say that for each $\g_k(t)\neq0$ the corresponding $k$-th channel is active. 

First, we fix the time-dependent probability distribution and expose the Kraus operators $\sig_k$ to the noise (perturb them). Having defined the probability distribution $p_k(t)$ such that it stems from Markovian dynamics (semigroup and CP-divisible), we look at the Markovian character of the perturbed dynamics, whether it was preserved or lost. This approach seems to be justified, since in real-life experiments we have to take into account unknown noise that may or may not destroy important features of our evolution e.g. Markovianity or semigroup property.  

The paper is organised as follows. In the next section, we elaborate on the model of noisy dynamics. Then we introduce the transfer matrix notation and present an algorithm for checking Markovianity of a dynamical map. Next, we discuss the numerical results of perturbed dynamics (based on divisibility and fidelity of dynamical maps). In the last section, we summerize our results and leave some questions for the further investigation.   

\section{The model} 
The first case we investigate is based on the Markovian semigroup of single dephasing channel that is
\begin{equation}\label{ideal}
\La_t^{(k)}(\rho)=\frac{1}{2}\big(1+e^{-\g_k t}\big)\rho+\frac{1}{2}\big(1-e^{-\g_k t}\big)\sig_k\rho\sig_k^\dag.
\end{equation}
For $\g_k>0$ $\La_t$ is CPTP and defines Markovian semigroup property.  Here we call Eq. (\ref{ideal}) the ideal semigroup equation, because the Kraus operators $\sig_k$ are not perturbed. However, in real-life experiments quantum systems are subjected to the additional, not-known noise. This noise, may stem from non-ideal preparation of laboratory equipment such as: mirrors, wave-plates, fibre-optics or detectors. Therefore, we modify Eq. (\ref{ideal}) as 
\begin{equation}\label{noisy}
\tilde{\La}_t^{(k)}(\rho)=\frac{1}{2}(1+e^{-\g_k t})\rho+\frac{1}{2}(1-e^{-\g_k t})\tilde{\sig}_k\rho\tilde{\sig}_k^\dag,\quad \g_k>0,
\end{equation}
where $\tilde{\sig}_k$ ($k=1,2,3$) are non-ideal Pauli spin matrices (i.e. with an additional noise) \footnote{We do not perturb identity operator, i.e. $\ssig_0=\sig_0=\mathbbm{1}$.} and they are of the following form
\begin{eqnarray}
\ssig_x=\exp(i\alpha G_x),&\quad\mathrm{with}\quad& G_x=\left(\begin{array}{cc}1 & -1 \\-1 & 1\end{array}\right),\\
\ssig_y=\exp(i\beta G_y),&\quad\mathrm{with}\quad& G_y=\left(\begin{array}{cc}1 & i \\-i & 1\end{array}\right),\\
\ssig_z=\exp(i\omega G_z),&\quad\mathrm{with}\quad& G_z=\left(\begin{array}{cc}0 & 0 \\0 & 1\end{array}\right).\\
\end{eqnarray}
For $\alpha=\beta=\frac{\pi}{2}$, $\omega=\pi$  one recovers the standard Pauli matrices. It seems natural to introduce the noise in the exponents, because it corresponds to a non-ideal equipment. For example, one may interpret $\sig_x$ as an operator changing the polarisation of a photon by $\frac{\pi}{2}$, then $\ssig_x$ is an operator that changes polarisation by angle $\alpha$ (the real laboratory mirrors have $\al=\frac{\pi}{2}\pm\eps$, with small perturbation $\eps\ge0$). Here, we investigate Markovianity (in terms of divisibility and semigroup) for the generic angles $\alpha,\beta$ and $\omega$ in $[0,2\pi)$ of our  noisy dynamics (\ref{noisy}).

Further we check the robustness of Markovianity for more general dynamical maps that are governed by the random unitary dynamics \cite{fi1}
\begin{equation}\label{nrud}
\La_t(\rho)=p_0(t)\rho+\sum_{k=1}^3p_k(t)\ssig_k\rho \ssig_k^\dag,
\end{equation}
with
\begin{eqnarray}
p_0(t)&=&\frac{1}{4}\Big(1+e^{-2(\g_2+\g_3)t}+e^{-2(\g_1+\g_3)t}+e^{-2(\g_1+\g_2)t}\Big),\label{p0}\\
p_1(t)&=&\frac{1}{4}\Big(1+e^{-2(\g_2+\g_3)t}-e^{-2(\g_1+\g_3)t}-e^{-2(\g_1+\g_2)t}\Big),\\
p_2(t)&=&\frac{1}{4}\Big(1-e^{-2(\g_2+\g_3)t}+e^{-2(\g_1+\g_3)t}-e^{-2(\g_1+\g_2)t}\Big),\\
p_3(t)&=&\frac{1}{4}\Big(1-e^{-2(\g_2+\g_3)t}-e^{-2(\g_1+\g_3)t}+e^{-2(\g_1+\g_2)t}\Big).\label{p3}
\end{eqnarray}
For unperturbed Kraus operators (i.e. $\ssig_k=\sig_k$) Eqs. (\ref{p0}-\ref{p3}) define semigroup dynamics. Each non-zero $\g_k$ corresponds to an active dephasing channel. 

In our model we perturb only Pauli matrices, treating them as non-ideal laboratory equipment while keeping the probability distribution corresponding to Markovian semigroup fixed and check the Markovian (semigroup) character of the map. All our considerations are based on the numerical analysis.

\section{The transfer matrix algorithm} The transfer matrix approach to quantum dynamics is equivalent to the dynamical map approach and carries all information about the evolution of OQS, including the influence of the environment on the system of interest. It is mostly used in quantum channels description \cite{wolf2} and geometrical approach to Markovianity of OQS \cite{lorenzo}.

For a set of $n$-dimensional orthonormal operators in the Hilbert space $\{G_\al\}_{\al=1}^{n^2}$ ($\tr(G_\al G_\be^\dag)=\delta_{\al,\be}$) we may construct the transfer matrix that is isomorphic with a dynamical map $\La_t$
\begin{equation}
\Phi_{\al,\be}(t):=\tr\Big(G_\al^\dag\La_t(G_\be)\Big).
\end{equation}
Therefore, one represents the map $\La_t$ as
\begin{equation}
\La_t(\rho)=\sum_{\al,\be=1}^{n^2}\Phi_{\al,\be}(t)G_\al\tr(G_\be^\dag\rho).
\end{equation}
If one chooses {\em unit matrices} as a basis, i.e. $G_\al=\ket{k}\bra{l}$ (where $\al:=(k,l)$) then the transfer matrix $\Phi(t)$ relates to the Choi matrix \cite{choi, jam} $W(t)$ associated with the map $\La_t$ as
\begin{equation}\label{choi-fi}
W(t)=\frac{1}{n}\sum_{\al,\be=1}^{n^2}\Phi_{\al,\be}(t)G_\be\otimes G_\al,
\end{equation}
where $W(t)=(\mathbbm{1}\otimes\La_t)P_+$ ($P_+$ being maximally entangled state). A dynamical map $\La_t$ is CPTP if and only if the corresponding Choi matrix $W(t)$ is positive and $\tr W(t)=1$.  

The advantage of using the $\Phi$-matrix is that one may easily calculate relevant features of the dynamical map, in particular, divisibility of a map (i.e. Markovianity). It is easy to verify that a concatenation of two maps $\La_A$ and $\La_B$ results in ordinary matrix multiplication of the corresponding transfer matrices $\Phi_A$ and $\Phi_B$. Therefore, if the transfer matrix $\Phi(t,s)$ that is associated with a propagator $V_{t,s}$ ($\La_t=V_{t,s}\La_s$) yields positive Choi matrix $W(t,s)$ for all $t\ge s$ then  $\La_t$ is Markovian (i.e. $V_{t,s}$ is CP and thus $\La_t$ is CP-divisible). It is worth stressing that $\Phi(t,s)=\Phi(t)\Phi^{-1}(s)$, provided that $\Phi(t)$ is invertible for all $t\ge0$. 

To decide whether a given map $\La_t$ is Markovian or not, we use the transfer matrix algorithm that reads as follows
\begin{enumerate}
\item Construct the transfer matrix $\Phi(t)$ associated with $\La_t$ (use the {\em unit matrices} as a basis \footnote{In principal, an arbitrary basis may be used, however, {\em unit matrices} simplify further calculations.}).
\item Find the inverse $\Phi^{-1}$ and take it at the time $t=s$ \footnote{Provided that $\Phi^{-1}$ exists, otherwise we have non-divisible map}.
\item Construct the transfer matrix of a propagator $V_{t,s}$ as $\Phi(t)\Phi(s)^{-1}\equiv \Phi(t,s)$.
\item Associate the Choi matrix $W(t,s)$ with $\Phi(t,s)$ as in (\ref{choi-fi}).
\item Calculate the eigenvalues of $W(t,s)$.
\item Markovian (CP-divisible) evolution is for all positive eigenvalues of $W(t,s)$, if at least one is negative, then $\La_t$ is non-Markovian.
\end{enumerate}
This algorithm is applicable for both analytical and numerical calculations. However, the complexity is directly associated with the complexity of evaluating eigenvalues of an $n^2\times n^2$ matrix, which for larger systems may be challenging.

\section{Numerical approach}
\begin{figure}[h]
\includegraphics[scale=0.4]{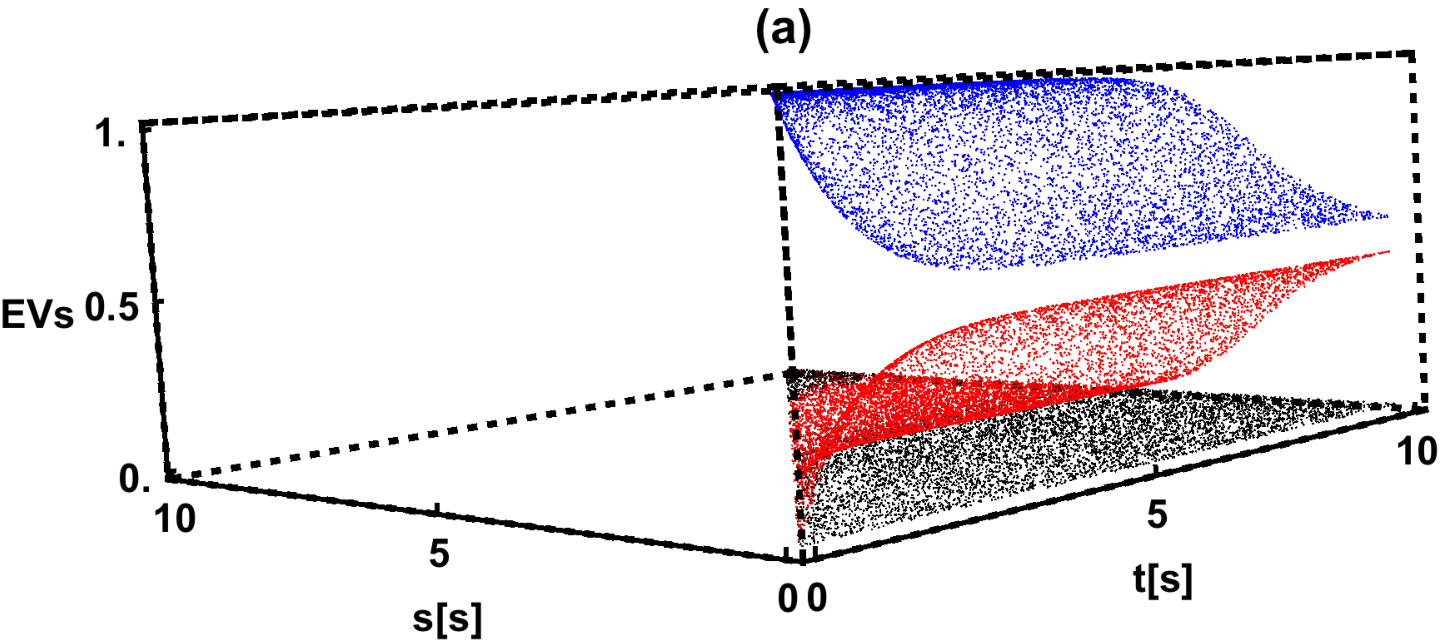}
\includegraphics[scale=0.4]{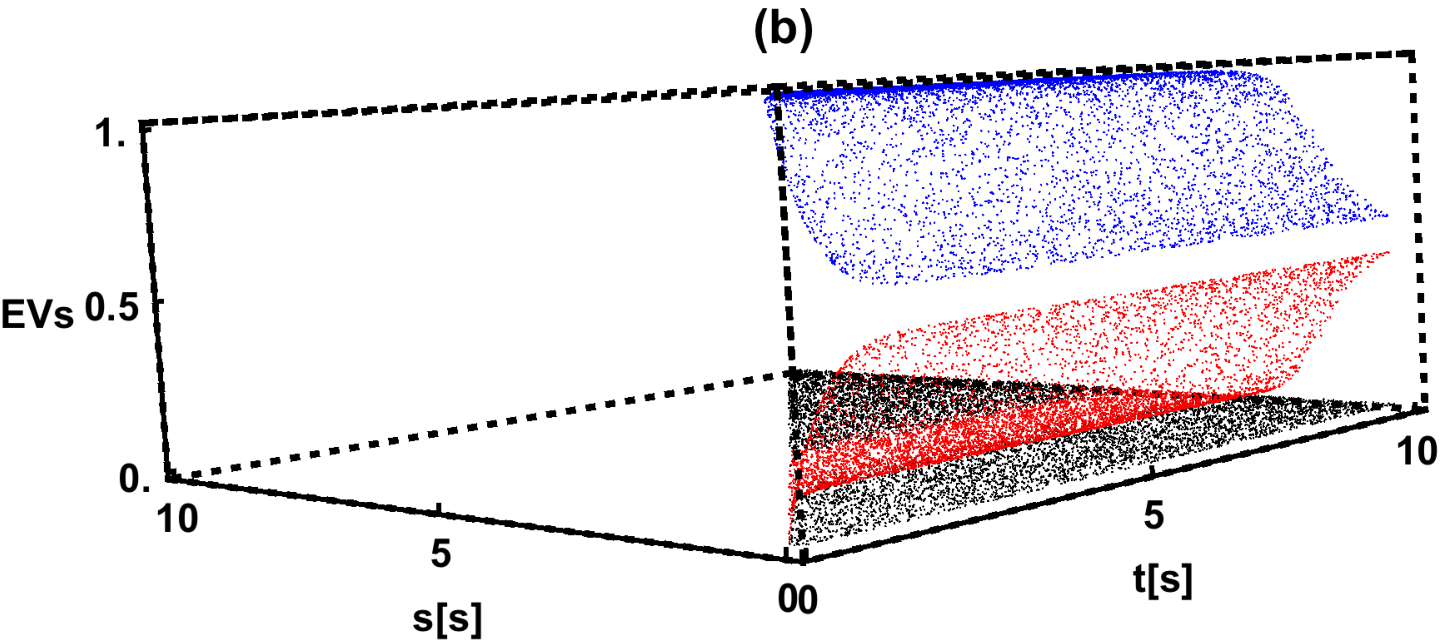}
\includegraphics[scale=0.4]{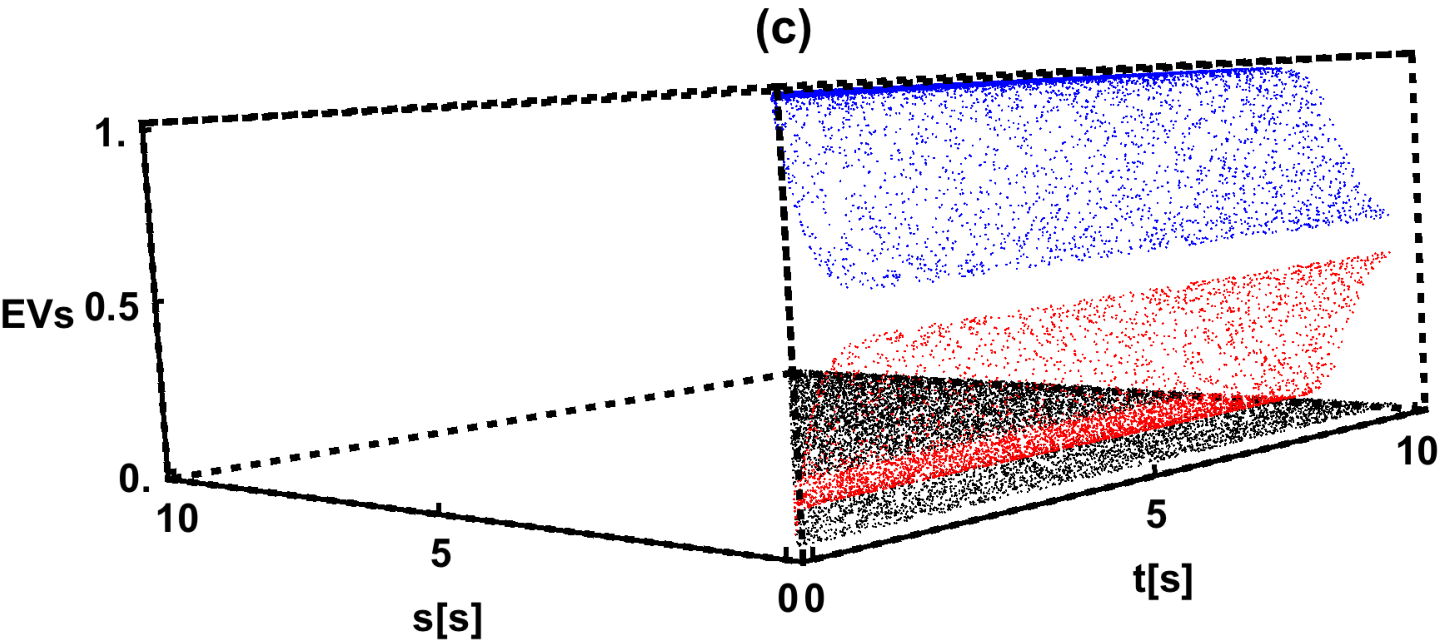}
\caption{Eigenvalues of Choi matrix associated with the propagator $\tilde{V}^{(x)}_{t,s}$ for a noisy dynamics $\tilde{\La}_t^{(x)}$. Different colors represent different eigenvalues (two eigenvalues are equal to zero). Each eigenvalues is composed of 10000 time points ($t>s\in[0,10]$). The noise of Kraus operator $\ssig_1$ is chosen as $\al=\frac{\pi}{2}+0.1$ and decoherence rates are $\g_1=1$ for (a), $\g_1=2$ for figure (b) and $\g_1=3$ for the (c).\label{evfig}}
\end{figure}

Let us consider the noisy single dephasing channel (\ref{noisy}). We apply the transfer matrix algorithm and use the numerical analysis for checking the Markovianity of $\La_t$. The parameters are selected as follow
\begin{itemize}
\item $\alpha, \beta, \omega\in [0,2\pi]$ - quantifying the noise in the Kraus operators (angles),
\item $\g_1,\g_2,\g_3\in[0,5]$ - decoherence rates for a Markovian semigroup dynamics \footnote{For numerical purposes we choose the same decoherence rate for all channels},
\item $t,s\in[0, 5]$ - time steps ($t> s)$ \footnote{For $\g_k\in[0,1]$ we choose time interval $[0,10]s$, while for rest we keep time up to 5s. We claim that this is sufficient for our purposes since all time-dependent probabilities exhibit exponential decay and larger times become negligible. This strong exponential dependance may be observed in the plots of eigenvalues.}.  
\end{itemize}
We perform the calculations for 100 choices of parameter $\g_k$, for which we form 1000 different noise Kraus operators and evaluate for each of them 5000 time steps ($t>s$). We observed that within numerical precision ($\sim10^{-9}$) the Markovianity of  $\La_t$ is preserved, while the semigroup property is lost. This suggests that the semigroup property is extremely fragile for adding uncontrolled perturbation (in our case perturbations in Kraus operators). On the other hand, Markovianity seems to be more robust. The eigenvalues of $\tilde{W}^{(x)}(t,s)$ for $\alpha=\frac{\pi}{2}+0.1$ and $\g_1=1,2,3$ are depicted in Fig. \ref{evfig}, that presents 10000 time points ($t>s\in[0,10]$).

\begin{figure}
\includegraphics[scale=0.5]{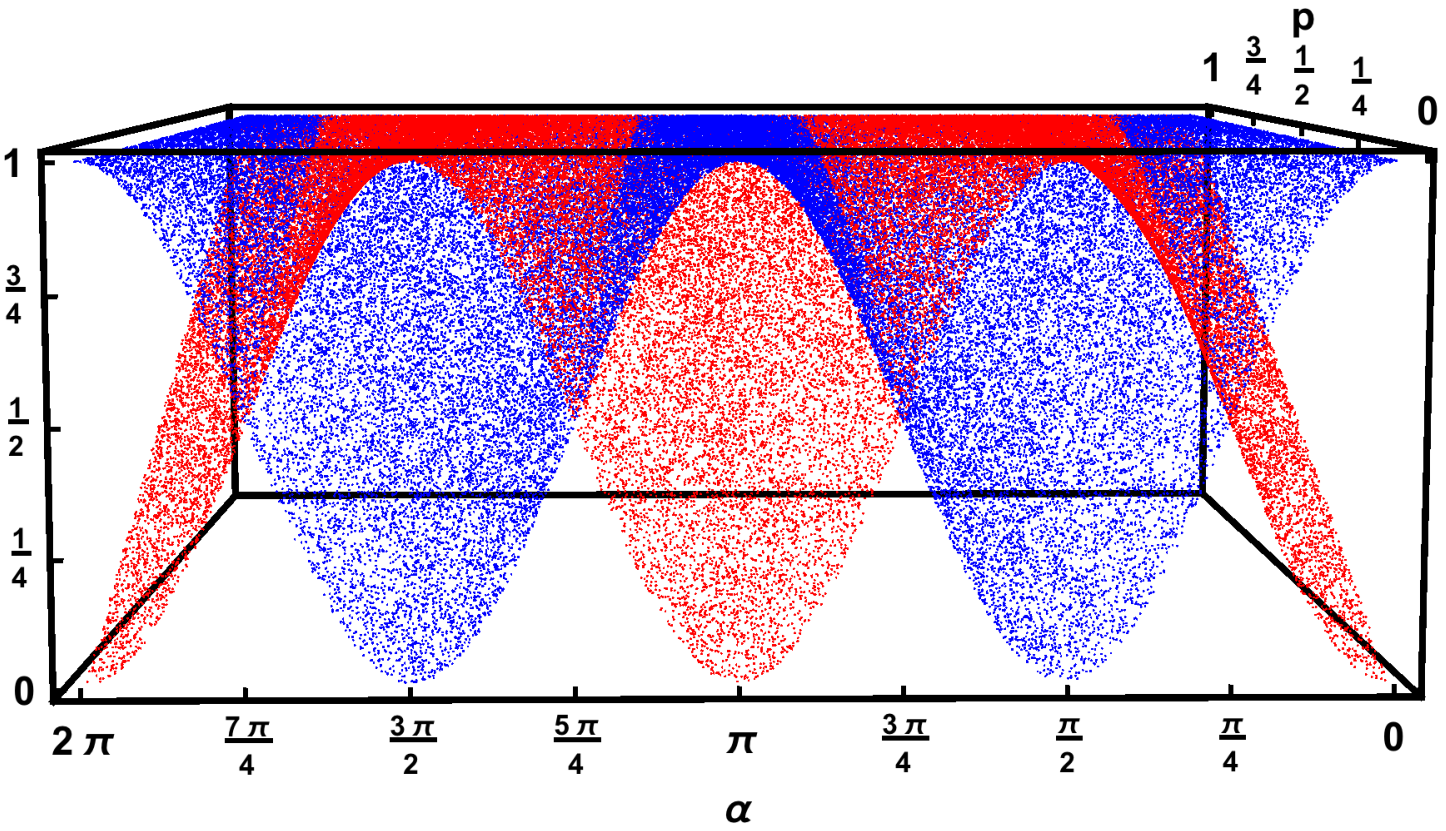}
\includegraphics[scale=0.5]{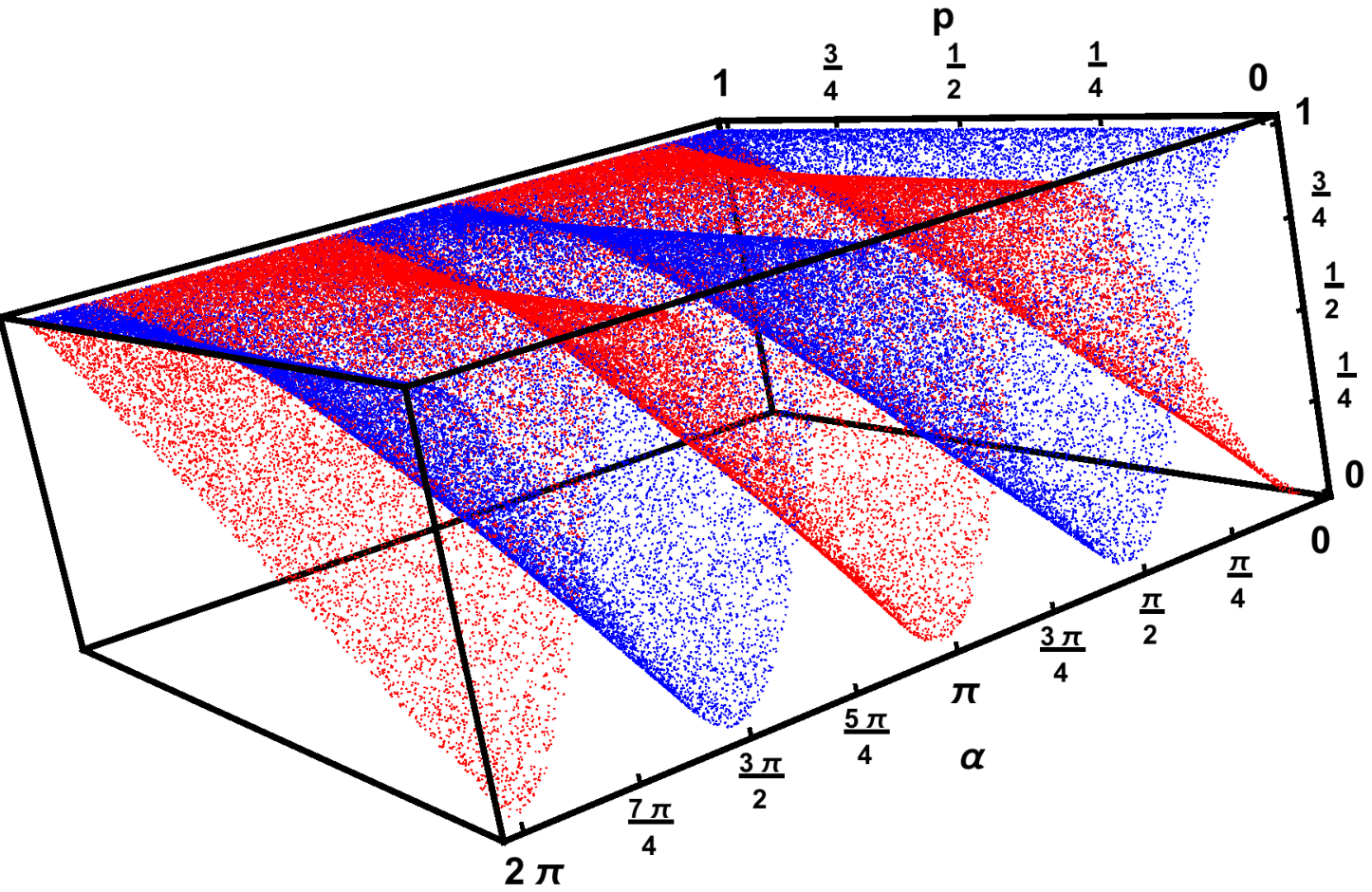}
\caption{Fidelity $F$ between Choi matrices of ideal (unperturbed) map and noisy one (the blue) and between noisy map and maximally entangled state (i.e. identity channel) (the red). All maps are single dephasing channels along $x$-axis. Parameter $p\in[0,1]$ represents map at ``different'' time steps, while $\al\in[0,2\pi]$ quantifies the noise in Kraus operator. The plot is depicted from two perspectives, from which the periodic character in noise angle is visible, and monotonic behavior in $p$ can be observed.\label{fid-tot} }
\end{figure}

One may wonder, how distant the ideal dynamics is from the noisy one? We provide this answer with the aid of a channel fidelity $F$ \cite{fid-chan, fid-package} for which it suffices checking the state fidelity \cite{Nilsen} between Choi matrices associated with the dynamical maps \footnote{This may be easily expressed in terms of Bures distance $D(\rho_1,\rho_2)=\sqrt{2(1-\sqrt{F(\rho_1,\rho_2)}}$.}. We perform this in two scenarios. The first one is based on the dynamical maps 
\begin{equation}
\La_t^{(x)}(\rho)=p \rho+(1-p)\sig_1\rho\sig_1,\quad \tilde{\La}_t^{(x)}(\rho)=p \rho+(1-p)\ssig_1\rho\ssig_1^\dag,
\end{equation}
where we change the parameters $p\in[0,1]$ \footnote{Note, that for semigroup $p\ge\frac{1}{2}$. It is important, that we look at fidelity between maps at a certain point $p\in[0,1]$, this approach cannot be directly associated with Markovianity. However, it gives some insight on the perturbation procedure. } and $\al\in[0,2\pi)$ (the noise parameter). The fidelity is computed between channel $\La_t^{(x)}$ and $\tilde{\La}_t^{(x)}$ as well as between identity channel (i.e. with maximally entangled state) and $\tilde{\La}_t^{(x)}$ (see Fig. \ref{fid-tot}). 

\begin{figure}
\includegraphics[scale=0.5]{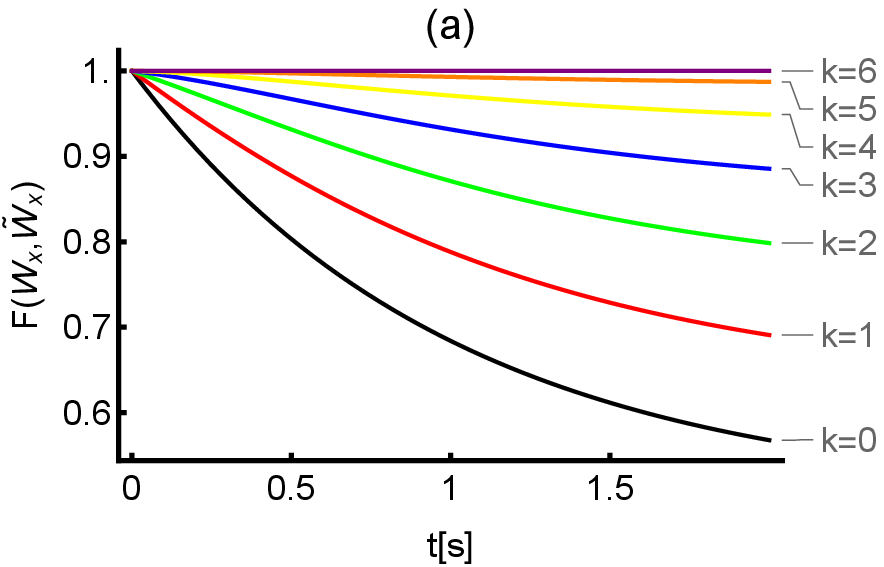}
\includegraphics[scale=0.5]{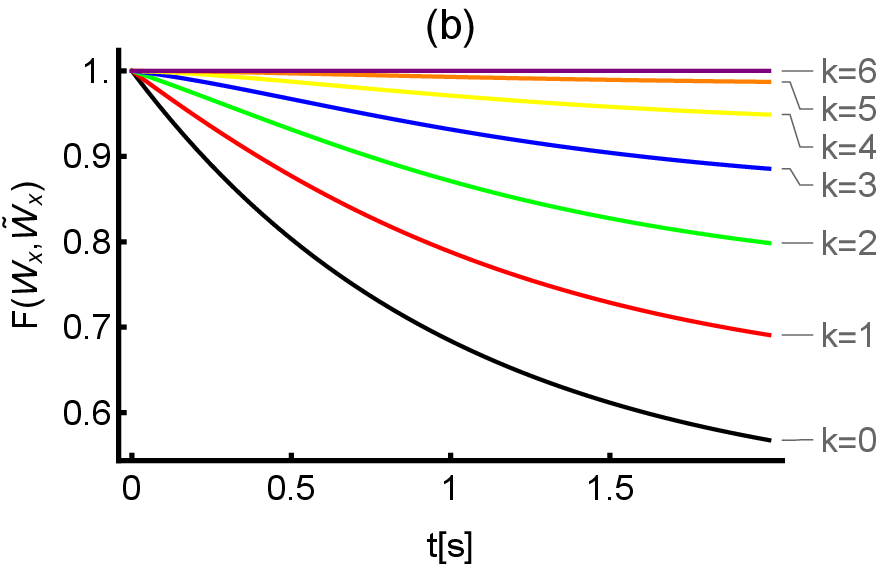}
\includegraphics[scale=0.5]{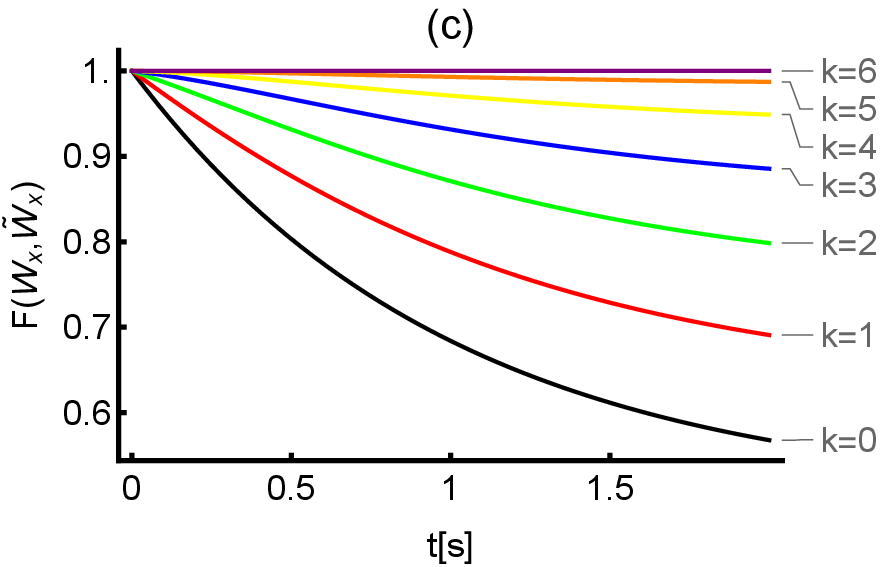}
\includegraphics[scale=0.5]{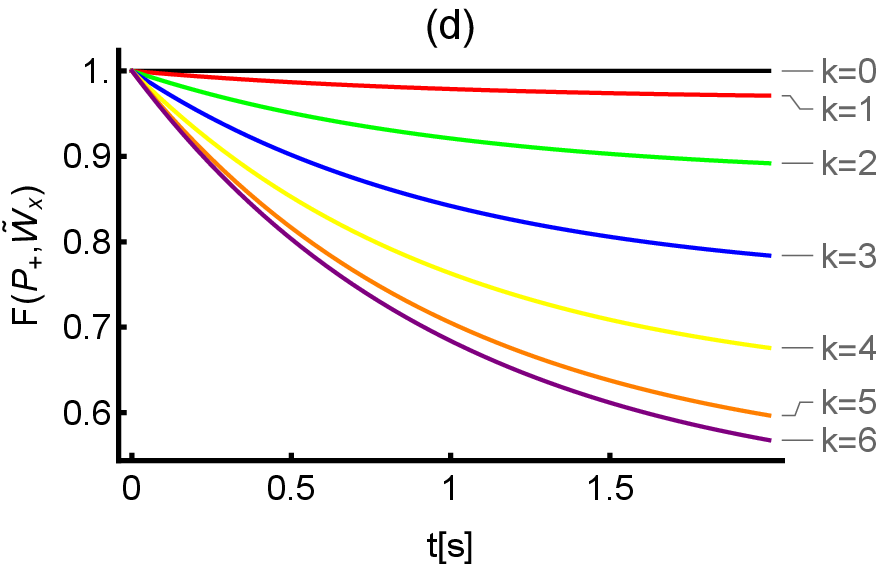}
\includegraphics[scale=0.5]{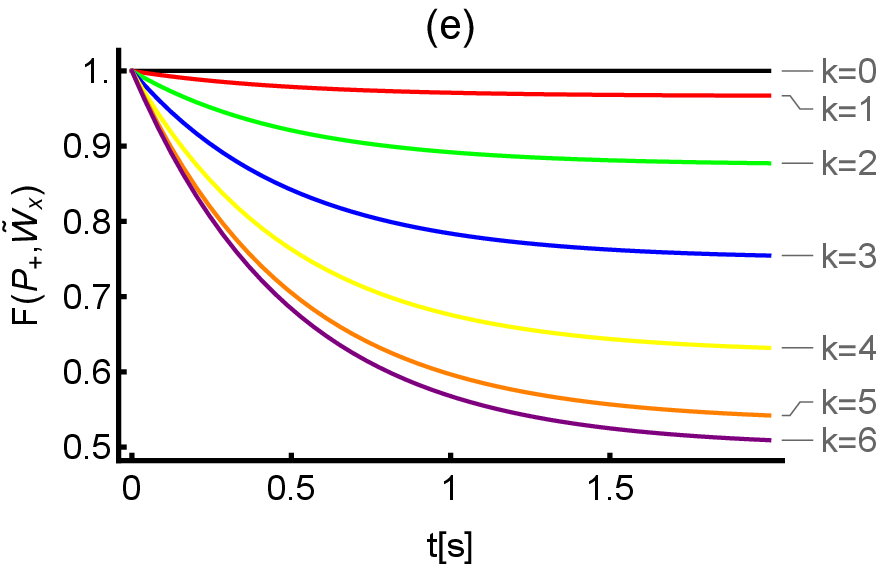}
\includegraphics[scale=0.5]{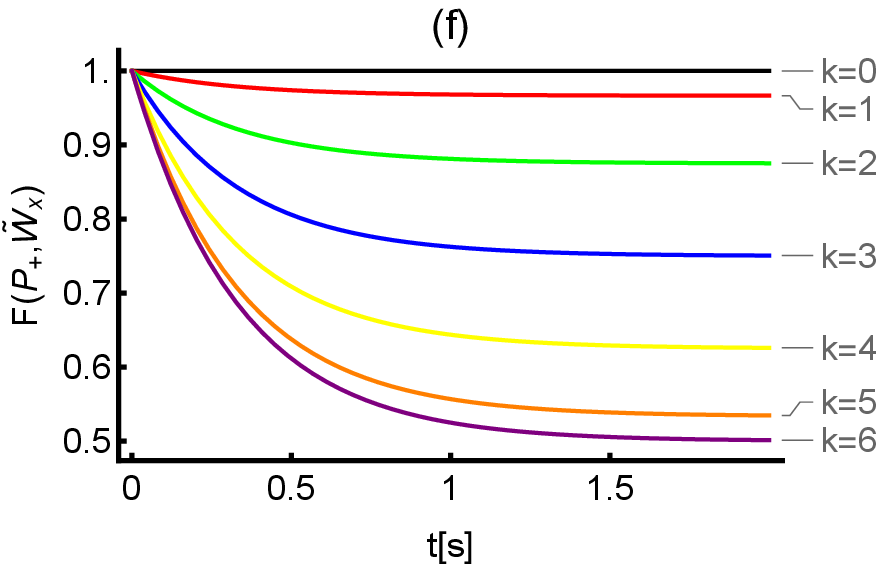}
\caption{Fidelity $F$ between Choi matrices of noisy dephasing channel along $x$-axis $\tilde{W}_x$ with the angles $\al=k\frac{\pi}{12}$ $(k=0,1,\ldots,6)$ and ideal semigroup $W_x$ (a-c).  Decoherence parameters are (a): $\g_1=1$, (b): $\g_1=2$ and (c): $\g_3=3$. Plots (d-f) depict fidelity between noisy channel and identity channel for the  angles $\al=k\frac{pi}{12}$ $(k=0,1,\ldots,6)$ and set of parameters (d): $\g_1=1$, (e): $\g_1=2$ and (f): $\g_3=3$. \label{fid-k}}
\end{figure}

Secondly, we look at fidelity for $\g_1=1,2,3$ for which $\al=k\frac{\pi}{12}$ ($k=0,1,\ldots,6)$ (see Fig. \ref{fid-k}). Because in that scenario, fidelity is a periodical function, we may limit ourselves up to $k=6$ instead of $k=24$. 

Both approaches to fidelity (the $y$ and $z$ dephasing channels behave symmetrically) show that as we move away from $\al=\frac{\pi}{2}$ towards $\al=0$ or $\al=\pi$ the fidelity between $\La_t$ and $\tilde{\La}_t$ decreases, achieving its minimum at $\al=0,\pi$. Decrease (monotonic) is also observed as a function of time (or equivalently as $p$ goes to zero). Now, if we compare $F(W(t),\tilde{W}(t))$ with $F(P_+,\tilde{W}(t))$ one notices that as channels move further (in $\al$ parameter) from the ideal semigroup they come closer to identity channel and vice versa. 

\begin{figure}
\includegraphics[scale=0.5]{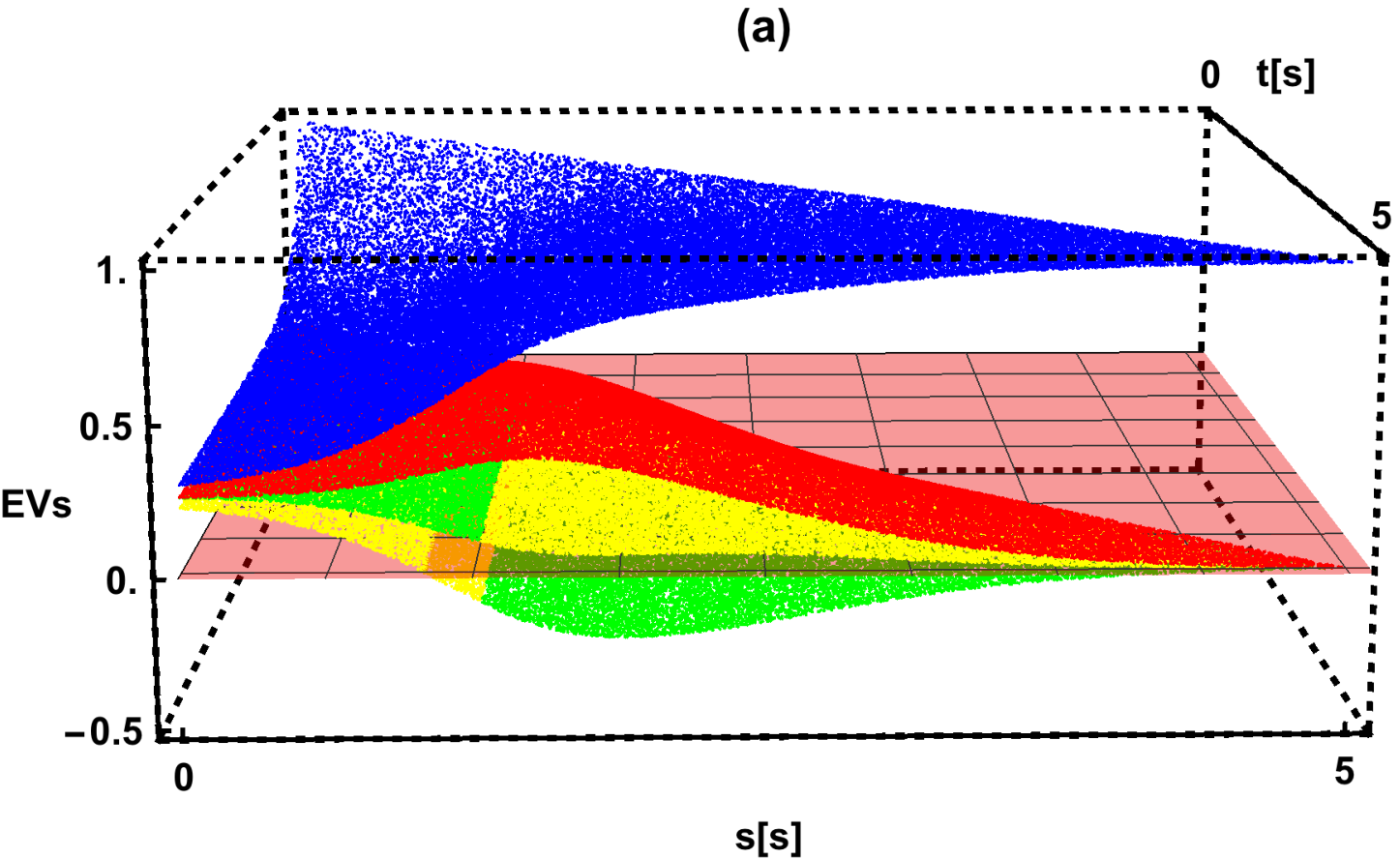}
\includegraphics[scale=0.5]{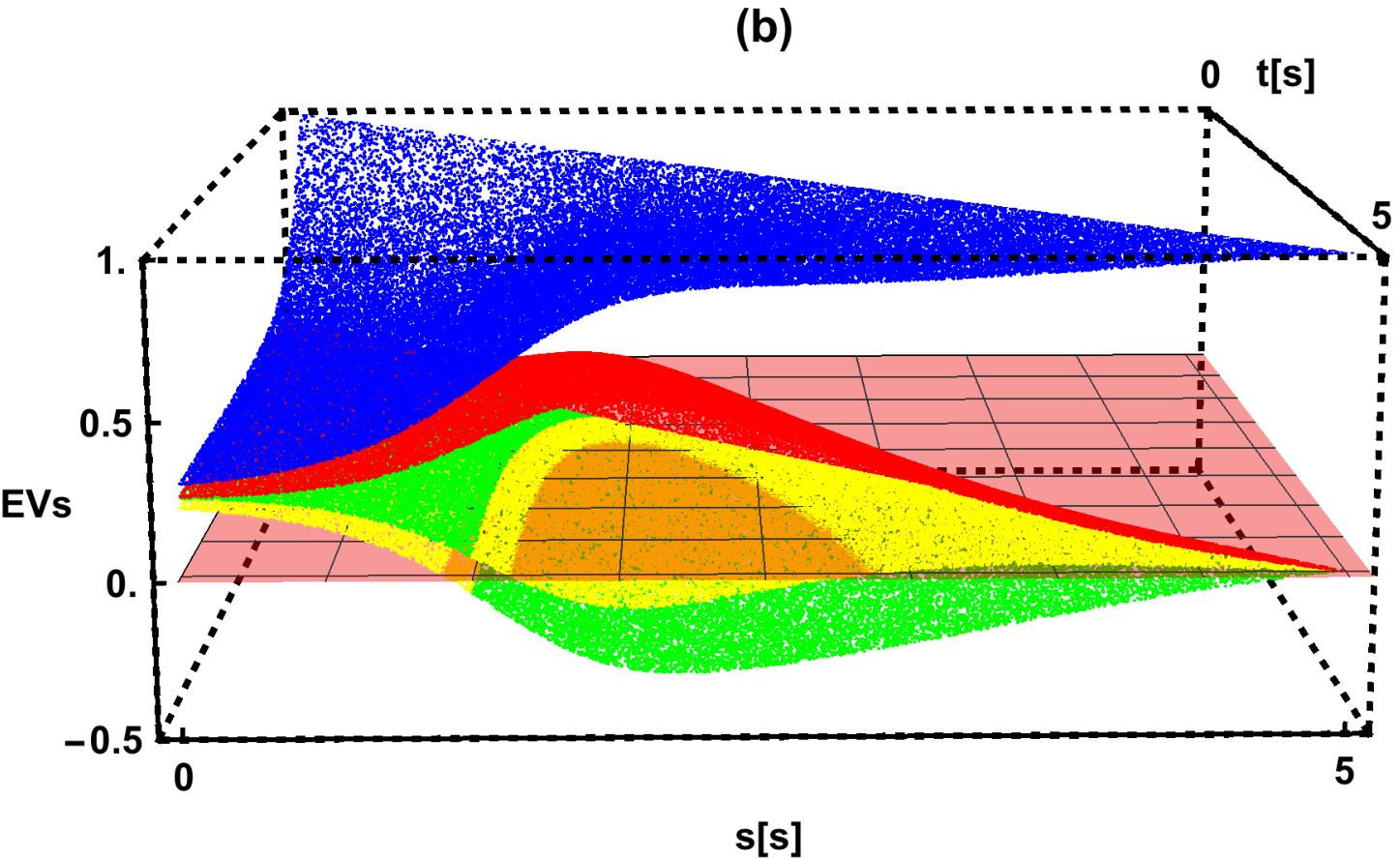}
\caption{Eigenvalues of the Choi matrix for a propagator $V_{t,s}$ of a noisy semigroup (\ref{nrud}) with noise angles $\al=\frac{\pi}{2}+0.1,\beta=\frac{\pi}{2}+0.1$ and $\omega=\pi+0.1$ and decoherence rates (a): $\g_1=1, \g_2=\frac{1}{2}, \g_3=\frac{3}{2}$ (all channels are active) and (b): $\g_1=1, \g_2=0, \g_3=\frac{3}{2}$ ($\sig_y$ channel off). Plane represents zero value eigenvalue, thus all points beneath manifest non-Markovianity (violation of CP-divisibility). \label{evs-nrud} }
\end{figure}

The robustness of Markovianity (CP-divisibility) for a single dephasing channel may suggest that noise transformations preserve the CP-divisibility for a semigroup. However, further investigations show that this is not the case even for qubit dynamics (\ref{nrud}). If we consider two or three active channels (two or three $\g_k\neq0$ in (\ref{nrud})), then the dynamical map is no longer Markovian. Our claim is supported by the numerical analysis, which shows negative eigenvalue of the Choi matrix associated with the propagator $V_{t,s}$ of the map (\ref{nrud}). The perturbation violates Markovianity significantly (see Fig. \ref{evs-nrud}). Investigating this model we disprove the robustness conjecture in the most general form. However, it leaves room for a particular choice of noise and decoherence parameters that are robust under noisy transformation. 

\begin{figure}
\includegraphics[scale=0.4]{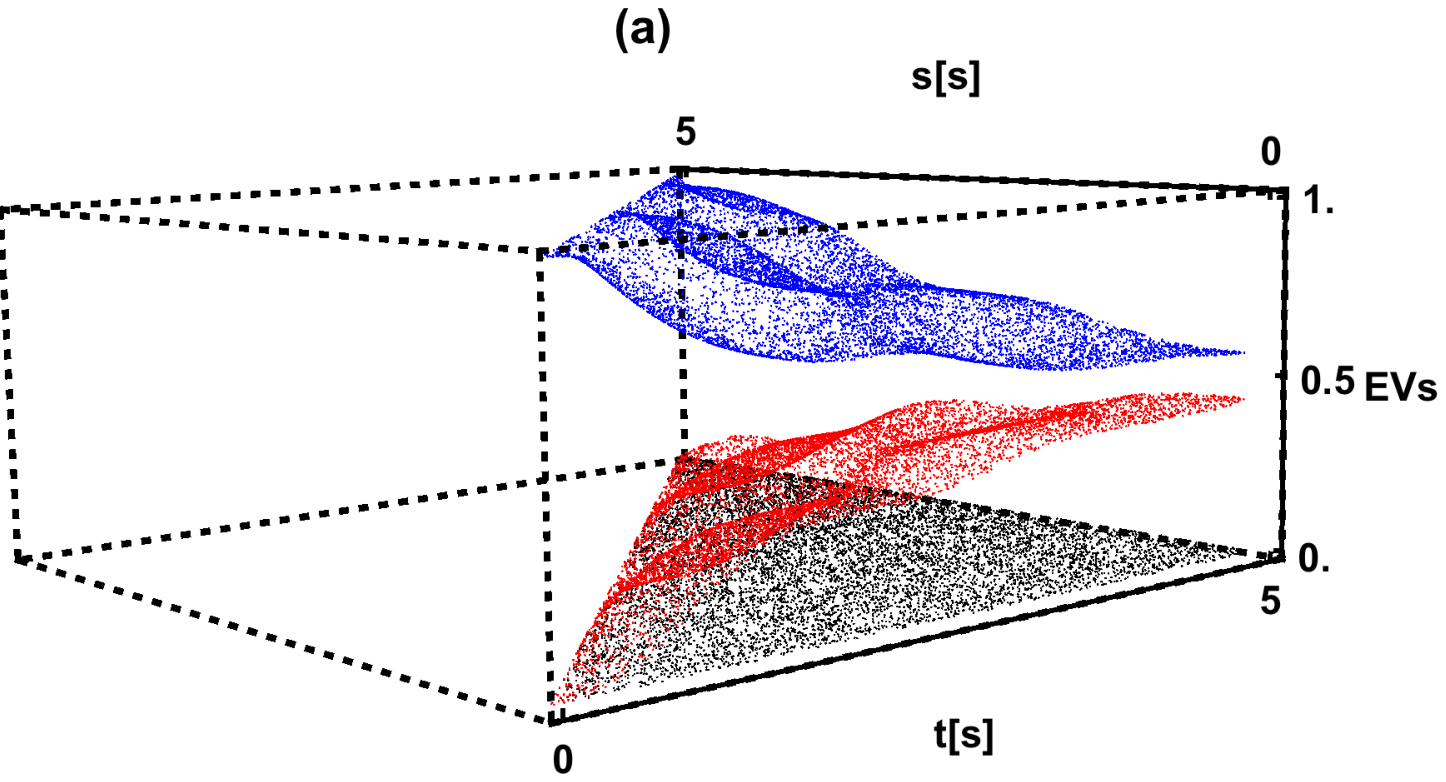}
\includegraphics[scale=0.4]{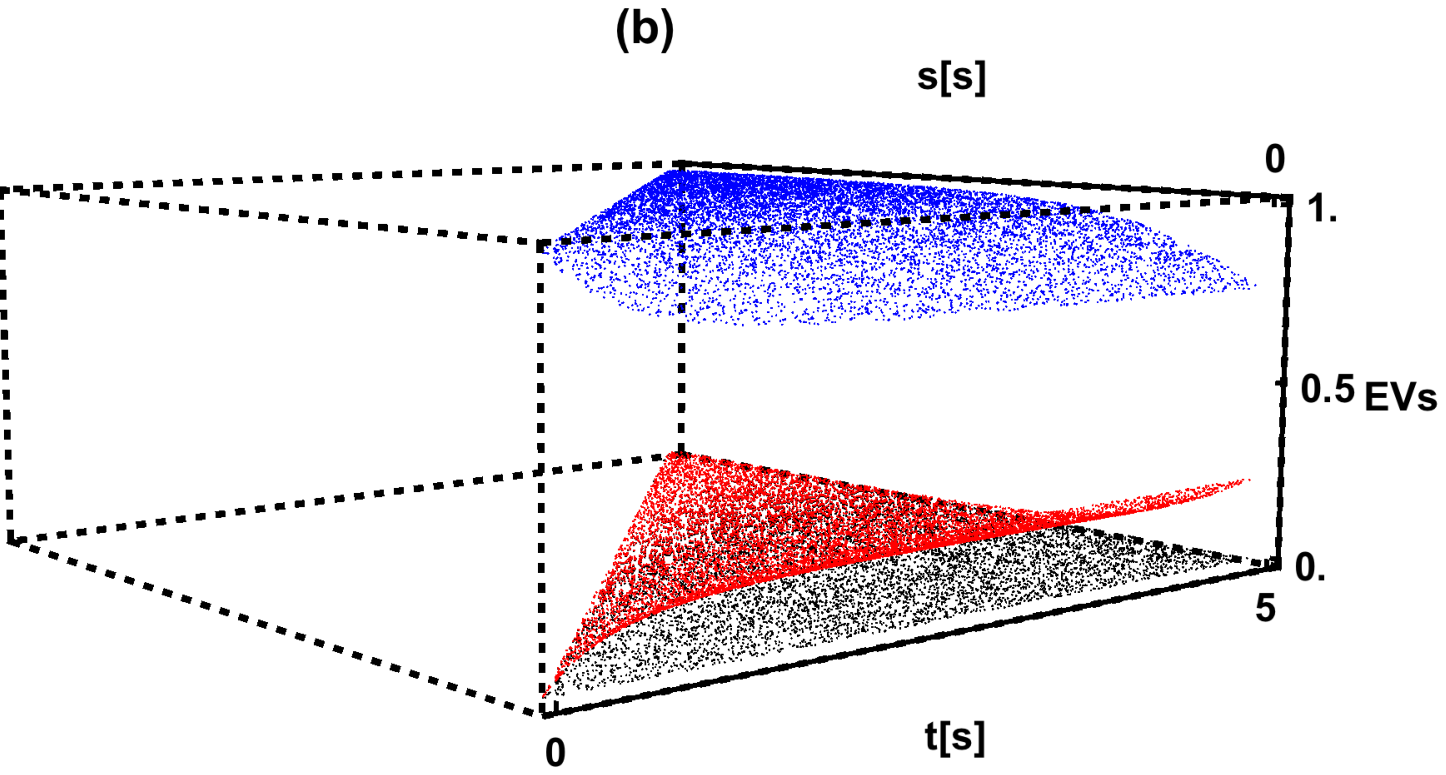}
\includegraphics[scale=0.4]{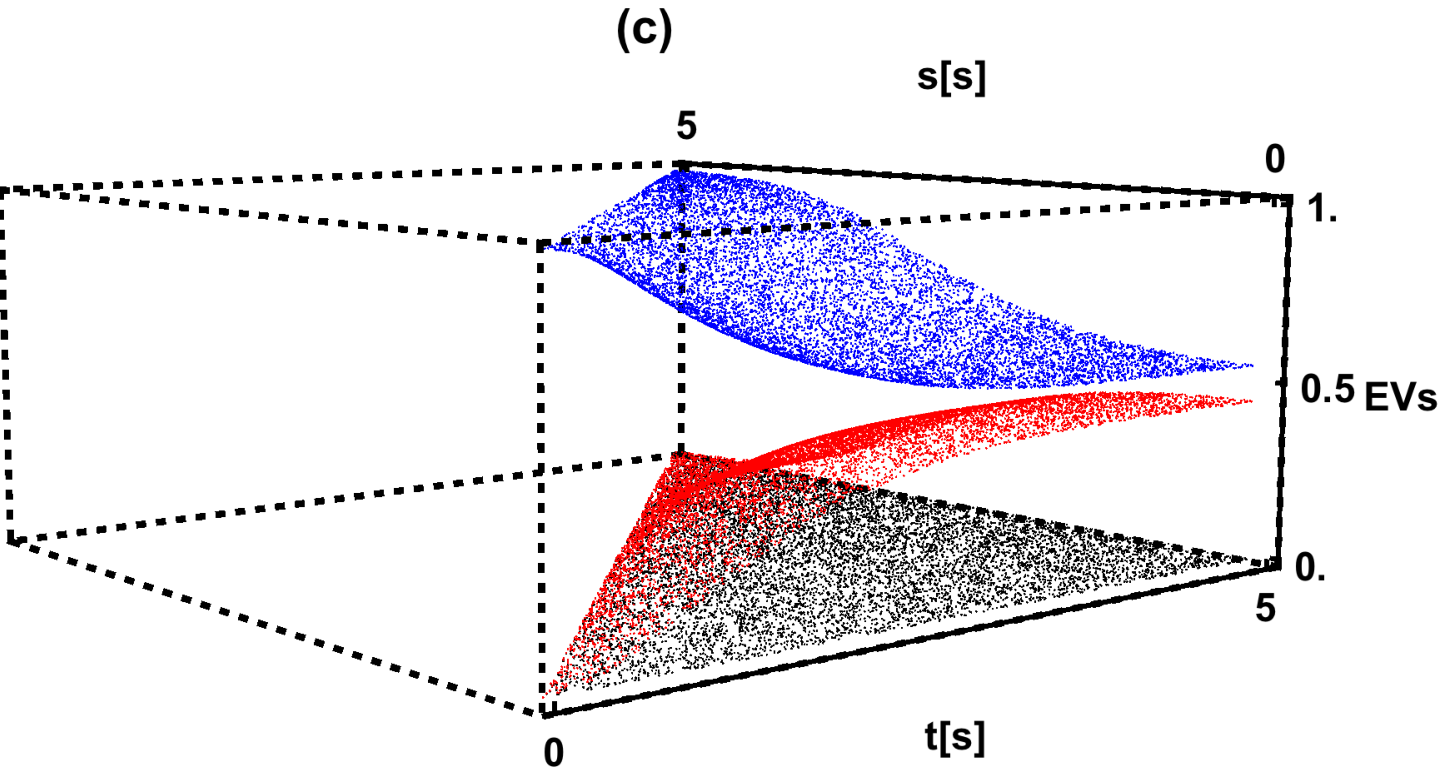}
\caption{Eigenvalues of the Choi matrix for a propagator $\tilde{V}^{(x)}_{t,s}$ of a noisy semigroup (\ref{nrud}) with the noise angle $\al=\frac{\pi}{2}+0.1$ and decoherence rates (a): $\g(t)=\sin^2(\frac{\pi}{2}t)$, (b): $\g(t)=e^{-t}$, (c): $\g(t)=\tanh(t)$.  \label{fig-td} }
\end{figure}

The same problem we may face with more general Markovian map, namely CP-divisible maps with time-dependent decoherence rates. It is well-known, that the dynamics is CP-divisible if and only if $\g_k(t)\ge0$ for all $t\ge0$. This gives us an additional degree of freedom, which is a positive function. We are unable to  check analytically Markovianity of a noisy dynamical map, as well as we cannot investigate all positive functions. Thus, we limited ourselves to the three types of positive functions $\sin^2(b t), e^{-a t}$ and $\tanh(a t)$. For each function, we chose 25 different $b\in(0,\pi)$ and $a\in(0,2)$, and for them we did the numerics of 1000 different noise operations and 5000 time steps $t>s$ for each perturbation. This was done for a single dephasing channel and we did not observe the violation of Markovianity. A particular choice of the parameters is depicted in the Fig. \ref{fig-td}. Therefore, we may pose a conjecture, that single dephasing Markovian channels (\ref{map}) are robust under noise transformation.

For at least two active dephasing channels (with time-dependent decoherence rates), the perturbation of Kraus operators destroys Markovian feature of the dynamical map. This is what one should expect, since semigroup is a special type of a positive function, namely a constant function. One may wonder, if there exist special types of function, that are robust under noise transformation. This, we leave as an open question.

\section{Conclusions}
We have shown how perturbation of Kraus operators influences the behavior of a dynamical map. Markovian semigroup single dephasing channels preserve Markovianity (CP-divisibility) and lose semigroup property.  If two or three dephasing channels are active, both semigroup and CP-divisibility are lost.  Our analysis was based solely on the numerical approach, for which we proposed an algorithm for checking the Markovianity of the evolution. This algorithm uses the notion of transfer matrix and shows its power in the field of quantum Markovianity. 

We have proposed a conjecture, that CP-divisiblity is also preserved for time-dependent decoherence rates in the scheme of single dephasing Markovian channels. In that case, we analysed three types of positive functions, that numerically indicate correctness of our statement. However, this requires further analysis.

One may wonder if the noise connected with non-ideal equipment appears in Kraus operators at the level of the dynamical map or in the operators of the generator. We leave this as an open problem for the future research.

\section{Acknowledgments}
This work is based upon research supported by the South African Research Chair Initiative of the Department of Science and Technology and National Research Foundation.

\end{document}